\documentclass[aps,pra,showpacs,twocolumn,superscriptaddress]{revtex4-1}
\usepackage{graphicx}
\usepackage[usenames]{color}
\usepackage{amssymb,amsmath}

\newcommand{\eq}[1]{Eq.~(\ref{#1})}
\newcommand{\fig}[1]{Fig.~\ref{#1}}
\newcommand{\be}[1]{\begin{equation}\label{#1}}
\newcommand{\ee}{\end{equation}}
\begin{document}

\title{ Non-dipole recollision-gated double ionization and observable effects }
\author{A. Emmanouilidou}
\address{Department of Physics and Astronomy, University College London, Gower Street, London WC1E 6BT, United Kingdom}
\author{T. Meltzer}
\address{Department of Physics and Astronomy, University College London, Gower Street, London WC1E 6BT, United Kingdom}
\author{P. B. Corkum}
\affiliation{Joint Laboratory for Attosecond Science, University of Ottawa and National Research Council,
100 Sussex Drive, Ottawa, Ontario, Canada K1A 0R6}
\begin{abstract}
Using a three-dimensional semiclassical model, we study  double ionization for strongly-driven He  fully accounting for magnetic field effects. For  linearly and  slightly elliptically polarized laser fields, we  show that  recollisions and  the magnetic field combined act as a gate. This gate favors more  transverse---with respect to the electric field---initial momenta of the tunneling electron   that are  opposite  to the   propagation direction of the laser field. In the absence of non-dipole effects, the transverse initial momentum is symmetric with respect to zero. We find that this asymmetry in the  transverse initial momentum  gives rise to an asymmetry in a double ionization observable. Finally,  we show that this asymmetry in the transverse initial momentum of the tunneling electron accounts for a recently-reported unexpectedly  large average sum of the electron momenta parallel to the propagation direction of the laser field.

\end{abstract}
\date{\today}

\maketitle

\section{Introduction}
Non-sequential  double ionization (NSDI) in driven two-electron atoms  is a  prototype process for exploring the electron-electron interaction in  systems driven by intense laser fields. As such, it  has attracted a lot of interest  \cite{NSDI1,NSDI2}.  Most theoretical studies on NSDI are formulated in the framework  of the dipole approximation where magnetic field effects are neglected  \cite{Becker1}. However, in the general case that the vector potential  $\mathrm{{\bf A}}$ depends on  both space and time, an electron experiences  a Lorentz force   whose magnetic field component is given by  $\mathrm{{\bf F_{B}}=q{\bf v}\times {\bf B}}$. We work in the non-relativistic limit. In this limit, magnetic-field effects   are expected to arise when the amplitude of the electron motion due to $\mathrm{{\bf F_{B}}}$  becomes 1 a.u., i.e.  $\mathrm{\beta_{0}\approx U_{p}/(2\omega c)\approx}$1 a.u.  \cite{Reiss1, Reiss2}. U$\mathrm{_p}$ is the ponderomotive energy. 
Non-dipole effects  were previously addressed in theoretical studies  of the observed ionization of Ne$\mathrm{^{n+}}$ ($\mathrm{n\leq8}$) in ultra-strong fields 
\cite{Walker1},  of stabilization \cite{add2} and of high-order harmonic generation  \cite{Joachain, Brabec,add1} as well as    in experimental studies \cite{Keller,Biegert}.   In recent studies in single ionization (SI), non-dipole effects of the electron momentum distribution along the propagation direction of the  laser field were addressed in experimental  \cite{Corkum1} and theoretical studies \cite{Corkum2,Corkum3,Drake, IvanovA}.

In this work, we show that in double ionization the magnetic field in conjunction with the recollision act as a gate.  This gate selects  a subset of the initial tunneling-electron momenta  along the propagation direction of the laser field. Only this subset leads to double ionization.  This gating is particularly pronounced at intensities smaller  than the intensities satisfying  the criterion for the onset of magnetic field effects $\mathrm{\beta_{0}\approx}$1 a.u.  \cite{Reiss1, Reiss2}. The propagation direction of the laser field is the same as the direction of the $\mathrm{\bf F_{B}}$ force (to first order). In  the current formulation, the  change in momentum due to the  $\mathrm{{\bf F_{B}}}$ force is along the +y-axis.  
The tunneling electron is the electron that initially tunnels in the field-lowered Coulomb potential. When non-dipole  effects are fully accounted for, we show that the y-component of the  initial momentum  of the tunneling-electron is mostly negative for events leading to double ionization. In the dipole approximation,  the initial momentum of the tunneling-electron that is transverse to the direction of the electric field is symmetric with respect to zero. The term  {\it non-dipole recollision-gated  ionization} is adopted to describe  ionization resulting from an asymmetric distribution of the   transverse tunneling-electron initial momentum due to the combined effect of the recollision and the magnetic field.  {\it Non-dipole recollision-gated  ionization} is a general phenomenon. We find that it underlies double electron escape in atoms driven by linearly and slightly elliptically polarized laser fields.

Moreover, we show that  {\it non-dipole recollision-gated  ionization}   results  in an asymmetry in a double ionization observable. Let $\mathrm{\phi\in [0^{\circ},180^{\circ}]}$ denote the angle of the final  ($\mathrm{t\rightarrow \infty}$) momentum of each escaping electron  with respect to 
 the propagation axis of the laser field. The observable in question is  P$\mathrm{_{asym}^{DI}(\phi)=}$P$\mathrm{^{DI}(\phi})$-P$\mathrm{^{DI}(180^{\circ}-\phi})$, 
 where P$\mathrm{^{DI}(\phi})$ is the  probability of either one of the two  electrons to escape with an angle $\mathrm{\phi}$.
   P$\mathrm{^{DI}(\phi})$ and P$\mathrm{_{asym}^{DI}(\phi)}$ are accessible  by  kinematically complete experiments.  In the dipole approximation,  P$\mathrm{_{asym}^{DI}(\phi)}=0$. When non-dipole effects are accounted for, it is shown that
 P$\mathrm{_{asym}^{DI}(\phi)}>0$,  for $\mathrm{\phi\in [0^{\circ},90^{\circ}]}$. This is in accord with  the effect of $\mathrm{\bf F_{B}}$. We also find that P$\mathrm{_{asym}^{DI}(\phi)}$ has considerable values over a wide interval  of $\mathrm{\phi}$   at lower intensities. This latter feature is an unexpected one. For the intensities considered the  $\mathrm{\bf F_{B}}$ force has small magnitude that increases with intensity. Thus,  
one would expect  the distribution P$\mathrm{_{asym}^{DI}(\phi)}$ to be very narrowly peaked around $\mathrm{90^{\circ}}$ with values increasing with intensity.

We finally show that   {\it non-dipole recollision-gated  ionization} is the mechanism underlying the surprisingly large average sum of the   momenta of the two escaping electrons along the propagation direction of the laser field. This large average sum of the electron momenta  is roughly an order of magnitude larger than twice the  average of the respective electron momentum in single ionization. We recently reported this in     ref.\cite{Emmanouilidoumagnetic1} for intensities around 10$^{15}$ Wcm$^{-2}$ for He at 800 nm (near-infrared) and around 10$^{14}$ Wcm$^{-2}$ for Xe at 3100 nm (mid-infrared)  \cite{Emmanouilidoumagnetic1}.      If magnetic-field effects are not accounted for the average momentum along the propagation direction of the laser field is zero.

 \section{Model}
We study ionization in strongly-driven He using a  three-dimensional (3D) semiclassical model that fully accounts for the magnetic field during time propagation---3D-SMND model.  We developed this model in ref.\cite{Emmanouilidoumagnetic1} by   extending a previously developed  3D semiclassical model in the framework of the dipole approximation---3D-SMD model \cite{Agapi1,Agapi2}. The Hamiltonian describing the interaction of the fixed nucleus two-electron atom with the laser field    is given by
\begin{equation}
\begin{aligned}
\mathrm{H=}&\mathrm{\frac{({\bf p}_{1}+{\bf A}(\mathrm{y_{1},t)})^2}{2}+\frac{({\bf p}_{2}+{\bf A}(\mathrm{y_{2},t}))^2}{2}-}\\
&\mathrm{-c_{1}\frac{Z}{|{\bf r}_{1}|}-c_{2}\frac{Z}{|{\bf r}_{2}|}+c_{3}\frac{1}{|{\bf r}_{1}-{\bf r}_{2}|}}.
\end{aligned}
\label{eqn:1}
\end{equation}
$\mathrm{{\bf A}}$ is the vector potential  given by
\begin{equation}
\mathrm{{\bf A}(y,t)=-\frac{E_{0}}{\omega}e^{-(\frac{ct-y}{c\tau})^2}(\sin{(\omega t-ky) \hat{x}}+\chi \cos (\omega t - k y)\hat{z})},
\label{eqn:2}
\end{equation} 
$\mathrm{\omega}$, k,  E$_{0}$ are the frequency, wavenumber, and strength of the electric component of the laser field,  respectively, and $\mathrm{\chi}$ is the ellipticity. c is the velocity of light and $\mathrm{\tau=FWHM/\sqrt{ln4}}$ with FWHM the full-width-half-maximum  of the laser field. All Coulomb forces are accounted for by setting $\mathrm{c_{1}=c_{2}=c_{3}=1}$. The  laser fields considered in the current work are either  linearly polarized, $\mathrm{\chi=0}$, or have a small ellipticity of $\mathrm{\chi=0.05}$.  For $\mathrm{\bf A}$ given by  \eq{eqn:2},  $\mathrm{\bf E}$ and $\mathrm{\bf B}$  are along the x- and z-axis, respectively, with small components along the z- and x-axis, respectively, for  laser fields with  $\mathrm{\chi=0.05}$. The propagation direction of the laser field and the direction of $\mathrm{\bf F_{B}}$ are mainly along the y-axis.  Unless otherwise stated,  all Coulomb forces as well as the electric and the magnetic field are fully accounted for during time propagation. To switch off a Coulomb interaction, the appropriate constant is set equal to zero. For example, to switch off the interaction of electron 1 with the nucleus,  $\mathrm{c_{1}}$ is set equal to zero. Moreover,  we address the Coulomb singularity by using  regularized coordinates \cite{KS}  which were also employed in the 3D-SMD  model \cite{Agapi1,Agapi2}. 

The initial state in the 3D-SMND model  entails one electron tunneling through the field-lowered Coulomb potential. The electron tunnels  with a non-relativistic quantum tunneling rate given by the Ammosov-Delone-Krainov (ADK) formula \cite{A1,A2}.  
A non-relativistic ADK rate results in this Gaussian distribution  being centered around zero. In ref.\cite{Keitel}   non-dipole effects were accounted for in the ADK rate. It was shown  that the most probable initial transverse momentum ranges from 0.33 I$\mathrm{_{p}}$/c  to almost zero with increasing  $\mathrm{E_{0}/(2I_{p})^{3/2}}$, with I$\mathrm{_{p}}$ the ionization energy of the tunneling electron.  
In this work, the smallest intensity considered is  7$\times$10$^{14}$ Wcm$^{-2}$ for He. At this intensity, if non-dipole effects are accounted for in the ADK rate,  the transverse momentum of the tunneling electron is centered around   0.12 I$\mathrm{_{p}^{He}}$/c for He which is  7.9$\times$10$^{-4}$ a.u.  (I$\mathrm{_{p}^{He}=}$0.904 a.u.). In what follows, we neglect this very small asymmetry and describe the distribution of the initial transverse momentum  by a Gaussian distribution centered around zero \cite{A1,A2}. We do so in order to clearly illustrate an important phenomenon, i.e. {\it non-dipole recollision-gated ionization}.
In addition, we set the initial momentum along the direction of the electric field equal to zero.    The remaining electron is initially described by a microcanonical distribution \cite{Abrimes}. We denote the tunneling and bound electrons  by electrons 1 and 2, respectively. The 3D-SMND model  is described in more detail in ref.\cite{Emmanouilidoumagnetic1}. 

\section{Non-dipole recollision gated ionization}
 In this work, we discuss non-dipole effects in double ionization. We do so  in the context of He when driven by an 800 nm, 12 fs FWHM   laser field that is linearly polarized  at intensities 1.3$\times$10$^{15}$ Wcm$^{-2}$,  2$\times$10$^{15}$ Wcm$^{-2}$ and 3.8$\times$10$^{15}$ Wcm$^{-2}$ and that is slightly elliptically polarized  with $\mathrm{\chi=0.05}$ at 2$\times$10$^{15}$ Wcm$^{-2}$.
 At these intensities the ponderomotive energy $\mathrm{U_{p}=E_{0}^2/(4\omega^2)}$ is 2.86 a.u., 4.39 a.u. and 8.35 a.u., respectively. Thus, the maximum energy of electron 1, 3.17$\mathrm{U_{p}}$, is above the energy needed to ionize He$^{+}$.  \subsection{Magnetic field asymmetry  in P$\mathrm{_{i}^{DI}(\phi)}$ }

First, we show that the magnetic field causes an asymmetry in the double ionization probability of electron i to ionize with an angle $\mathrm{\phi}$, which is denoted by P$\mathrm{_{i}^{DI}(\phi)}$ with $\mathrm{i=1,2}$ for electrons 1 and 2.   P$\mathrm{^{SI}(\phi)}$ is the corresponding  probability in single ionization.
$\mathrm{\phi}$ is the angle of the final momentum of  electron i  with respect to  the propagation axis of the laser field, i.e. $\mathrm{\cos{\phi}=p^{i}\cdot \hat{y}/|p^{i}|}$. The y-component of the electron momentum  is  parallel to the propagation direction of the laser field and to   $\mathrm{\bf F_{B}}$.

\begin{figure} [ht]
\includegraphics[clip,height=0.46\textwidth]{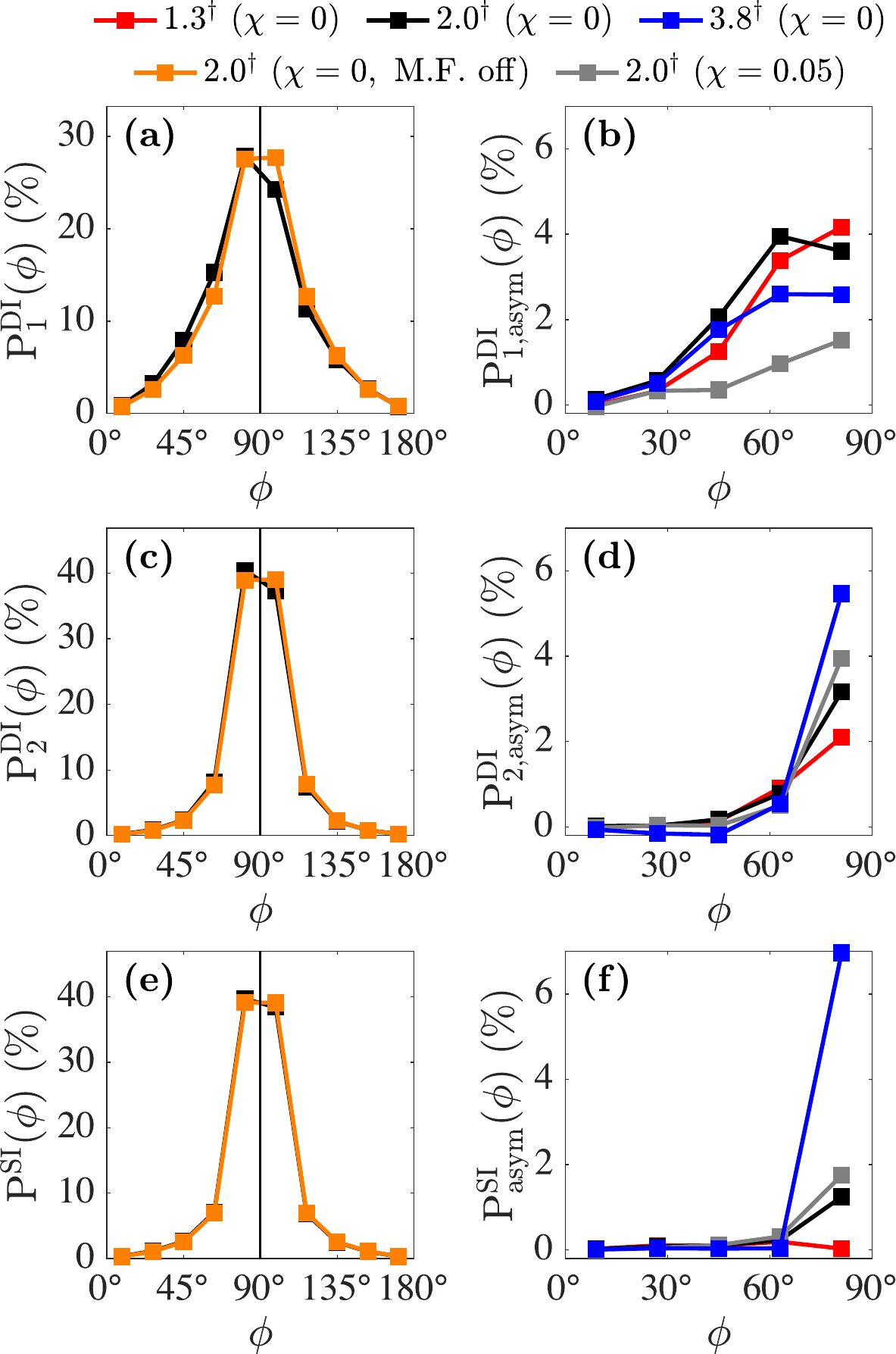}
\caption{ (a) P$\mathrm{_{1}^{DI}(\phi)}$ of  the tunneling electron, (c) P$\mathrm{_{2}^{DI}(\phi)}$ of the initially bound electron and (e) P$\mathrm{^{SI}(\phi)}$   in single ionization are plotted as a function of $\mathrm{\phi}$ at 2 $\times$10$^{15}$ Wcm$^{-2}$,   with the magnetic field  switched-on and off.  (b) P$\mathrm{_{1,asym}^{DI}(\phi)}$  of the tunneling electron, (d) P$\mathrm{_{2,asym}^{DI}(\phi)}$ of the initially bound electron  and  (f) P$\mathrm{_{asym}^{SI}(\phi)}$ in single ionization  are plotted as a function of $\mathrm{\phi}$ at three intensities with $\mathrm{\chi=0}$ and at one intensity with $\mathrm{\chi=0.05}$.
$\mathrm{\phi}$ is binned in intervals of 18$^{\circ}$ and  $\dagger$ denotes a multiplication factor of $\mathrm{10^{15}Wcm^{-2}}$. }
\label{figure4}
\end{figure}

  To show this asymmetry, we plot P$\mathrm{_{1}^{DI}(\phi)}$ of the tunnel electron and P$\mathrm{_{2}^{DI}(\phi)}$  of the initially bound electron   in \fig{figure4}(a) and (c), respectively, while we plot P$\mathrm{^{SI}(\phi)}$  in  \fig{figure4}(e). These plots are at an  intensity of  2$\times$10$^{15}$ Wcm$^{-2}$
  with the magnetic field switched-on and off. When the magnetic field is switched-on,  we find that all probability distributions  are asymmetric with respect to $\mathrm{\phi=90^{\circ}}$. 
This asymmetry is due to the magnetic field. Indeed,  when the magnetic field is switched-off all distributions 
 are shown to be symmetric with respect to $\mathrm{\phi=90^{\circ}}$. The latter is expected, since there is no preferred 
direction of electron escape on the plane that is  perpendicular to the x-axis (polarization direction). Moreover,  with the magnetic field switched-on, we find that P$\mathrm{_{i}^{DI}(\phi)>P_{i}^{DI}(180^{\circ}-\phi})$ and P$\mathrm{^{SI}(\phi)>P^{SI}(180^{\circ}-\phi})$ for  $\mathrm{\phi\in [0^{\circ},90^{\circ}]}$. Equivalently P$\mathrm{_{i,asym}^{DI}(\phi)=}$P$\mathrm{_{i}^{DI}(\phi})$-P$\mathrm{_{i}^{DI}(180^{\circ}-\phi})>0$ and P$\mathrm{_{asym}^{SI}(\phi)=}$P$\mathrm{^{SI}(\phi})$-P$\mathrm{^{SI}(180^{\circ}-\phi})>0$ for  $\mathrm{\phi\in [0^{\circ},90^{\circ}]}$. This is consistent with  the gain of momentum due to the $\mathrm{\bf F_{B}}$ force being along the +y-axis. That is, an electron is more likely to ionize with a positive rather than a negative y-component of the final momentum.

The asymmetry with respect to $\mathrm{\phi=90^{\circ}}$ is better illustrated   in \fig{figure4}. 
 We plot P$\mathrm{_{i,asym}^{DI}(\phi)}$ and P$\mathrm{_{asym}^{SI}(\phi)}$   as a function of $\mathrm{\phi}$   at  1.3$\times$10$^{15}$ Wcm$^{-2}$,  2$\times$10$^{15}$ Wcm$^{-2}$ and 3.8$\times$10$^{15}$ Wcm$^{-2}$ and at  2$\times$10$^{15}$ Wcm$^{-2}$ with $\mathrm{\chi=0.05}$.
Starting with single ionization,  $\mathrm{P_{asym}^{SI}(\phi)}$
 is almost zero at 1.3$\times$10$^{15}$ Wcm$^{-2}$. At the higher intensity of 3.8$\times$10$^{15}$ Wcm$^{-2}$,  $\mathrm{P_{asym}^{SI}(\phi)}$ is sharply centered around   90$^{\circ}$  reaching roughly 7\%, see \fig{figure4}(f). These features   of  P$\mathrm{^{SI}(\phi)}$ are in accord with the effect of the  $\mathrm{\bf F_{B}}$ force. $\mathrm{\bf F_{B}}$    is small for the intensities considered.  Therefore, $\mathrm{\bf F_{B}}$    has an observable effect mostly when the y-component of the electron  momentum is small as well, i.e. for an angle of escape $\mathrm{\phi=90^{\circ}}$. 
 In addition, $\mathrm{|\bf F_{B}|}$ is three times  larger for the higher intensity  compared to  the smaller one. As a  result   $\mathrm{P_{asym}^{SI}(\phi)}$ has larger values at higher intensities.

  In double ionization, we plot P$\mathrm{_{2,asym}^{DI}(\phi)}$ of  the initially bound electron  in \fig{figure4}(d). It is shown that 
 P$\mathrm{_{2,asym}^{DI}(\phi)}$ resembles mostly  P$\mathrm{^{SI}_{asym}(\phi)}$ rather than     P$\mathrm{_{1,asym}^{DI}(\phi)}$ in \fig{figure4}(b). Indeed, P$\mathrm{_{2,asym}^{DI}(\phi)}$ has larger values for higher intensities, as is the case for $\mathrm{P_{asym}^{SI}(\phi)}$, reaching roughly  5.5\% at 3.8$\times$10$^{15}$ Wcm$^{-2}$.
   We also find that the    distribution $\mathrm{P_{1,asym}^{DI}(\phi)}$ of the tunneling electron has different features from P$\mathrm{^{SI}_{asym}(\phi)}$, compare \fig{figure4}(b) with  \fig{figure4}(f).   We find that  $\mathrm{P_{1,asym}^{DI}(\phi)}$  is much wider than $\mathrm{P_{asym}^{SI}(\phi)}$.
  Also, for  $\mathrm{\phi\in [45^{\circ},90^{\circ}]}$, $\mathrm{P_{1,asym}^{DI}(\phi)}$   has higher values  at the smaller intensities of 1.3$\times$10$^{15}$ Wcm$^{-2}$ and 2$\times$10$^{15}$ Wcm$^{-2}$ rather than at  3.8$\times$10$^{15}$ Wcm$^{-2}$---4\% compared to 2.5\%.   
  We have  shown in ref.\cite{Emmanouilidoumagnetic1} that strong recollisions \cite{Corkum} prevail for strongly-driven He at 800 nm at  intensities of 1.3$\times$10$^{15}$ Wcm$^{-2}$ and 2$\times$10$^{15}$ Wcm$^{-2}$, while soft ones prevail at 3.8$\times$10$^{15}$ Wcm$^{-2}$. It then follows that
  $\mathrm{P_{1,asym}^{DI}(\phi)}$ has higher values for strong recollisions. This is also supported by the small values of $\mathrm{P_{1,asym}^{DI}(\phi)}$ at  2$\times$10$^{15}$ Wcm$^{-2}$ for a laser pulse with a small ellipticity of $\mathrm{\chi=0.05}$, see \fig{figure4}(b). We find (not shown)  that the recollisions are soft at  2$\times$10$^{15}$ Wcm$^{-2}$ for a laser pulse with $\mathrm{\chi=0.05}$. The times of recollision    correspond roughly to zeros of the laser field for strong recollisions and   extrema of the laser field for soft ones  \cite{Emmanouilidoumagnetic1,Agapi4}.  Moreover, the transfer of energy, compared to U$\mathrm{_p}$, from electron 1 to  electron 2 is larger for  a strong recollision and smaller for a soft one  \cite{Emmanouilidoumagnetic1,Agapi4}.  Later in the paper, we explain in  detail why the width and the values  of $\mathrm{P_{1,asym}^{DI}(\phi)}$ are large at 2$\times$10$^{15}$ Wcm$^{-2}$, smaller  at 3.8$\times$10$^{15}$ Wcm$^{-2}$ and even smaller at 2$\times$10$^{15}$ Wcm$^{-2}$ for a laser pulse with $\mathrm{\chi=0.05}$.

\begin{figure} [ht]
\includegraphics[clip,height=0.3\textwidth]{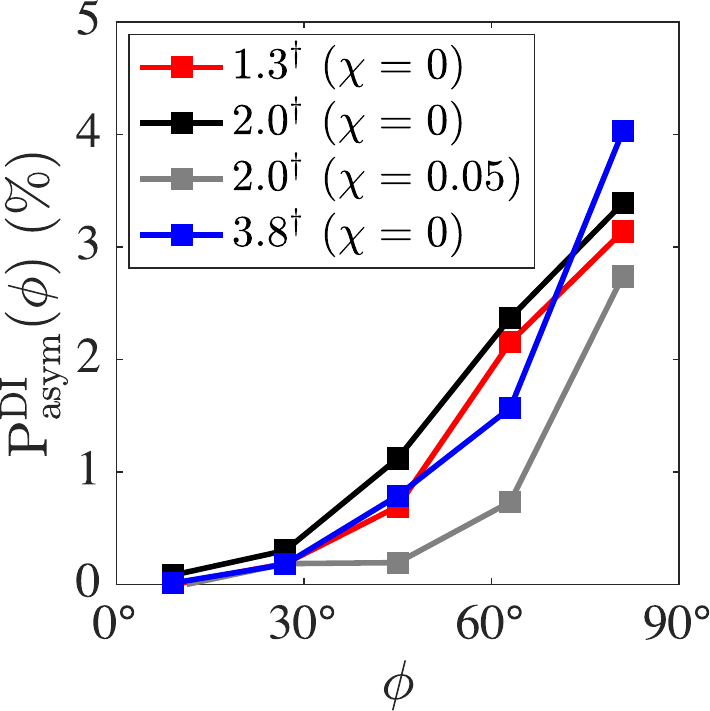}
\caption{In double ionization, P$\mathrm{_{asym}^{DI}(\phi)}$ for the two escaping  electrons is plotted as a function of $\mathrm{\phi}$.  Three intensities are considered and $\mathrm{\phi}$ is binned in intervals of 18$^{\circ}$.  $\dagger$ denotes a multiplication factor of $\mathrm{10^{15}Wcm^{-2}}$. } 
\label{figure9}
\end{figure}

Experimentally electron 1 can not be distinguished from electron 2. Therefore,  the probability distributions of electrons 1 and 2 to escape with an angle $\mathrm{\phi}$,  P$\mathrm{_{1}^{DI}(\phi)}$ (\fig{figure4}(a)) and  P$\mathrm{_{2}^{DI}(\phi)}$  (\fig{figure4}(c)), respectively, are not  experimentally
accessible. However, in a kinematically complete experiment, for each doubly-ionized event, the angle $\mathrm{\phi}$ of  each ionizing electron can be measured. Then, the probability distribution for any one of the two electrons to ionize 
with an angle $\mathrm{\phi}$, $\mathrm{P^{DI}(\phi)}$, can be obtained,  for $\mathrm{\phi\in [0^{\circ},180^{\circ}]}$. We compute and plot the distribution  $\mathrm{P_{asym}^{DI}(\phi)}$  as a function of $\mathrm{\phi}$ in \fig{figure9}. $\mathrm{P_{asym}^{DI}(\phi)}$
is found to have significant values at smaller intensities  over  the same wide range of $\mathrm{\phi}$ as  $\mathrm{P_{1,asym}^{DI}(\phi)}$ does. However, $\mathrm{P_{asym}^{DI}(\phi)<P_{1,asym}^{DI}(\phi)}$ at the smaller intensities. 
Moreover, at 3.8$\times$10$^{15}$ Wcm$^{-2}$ and at 2$\times$10$^{15}$ Wcm$^{-2}$ with $\mathrm{\chi=0.05}$, $\mathrm{P_{asym}^{DI}(\phi)>P_{1,asym}^{DI}(\phi)}$ at $\mathrm{\phi=81^{\circ}}$ but has non zero values for a wider range of $\mathrm{\phi}$ compared to  $\mathrm{P_{2,asym}^{DI}(\phi)}$.
 These features are expected since 
$\mathrm{P_{asym}^{DI}(\phi)}$ accounts for both the tunneling and the initially bound electron.   However, the features of the experimentally accessible  $\mathrm{P_{asym}^{DI}(\phi)}$ still capture the main features of $\mathrm{P_{1,asym}^{DI}(\phi)}$.

\subsection{Asymmetric transverse electron 1  momentum  at the tunnel  time }
\begin{figure} [ht]
\includegraphics[clip,width=0.5\textwidth]{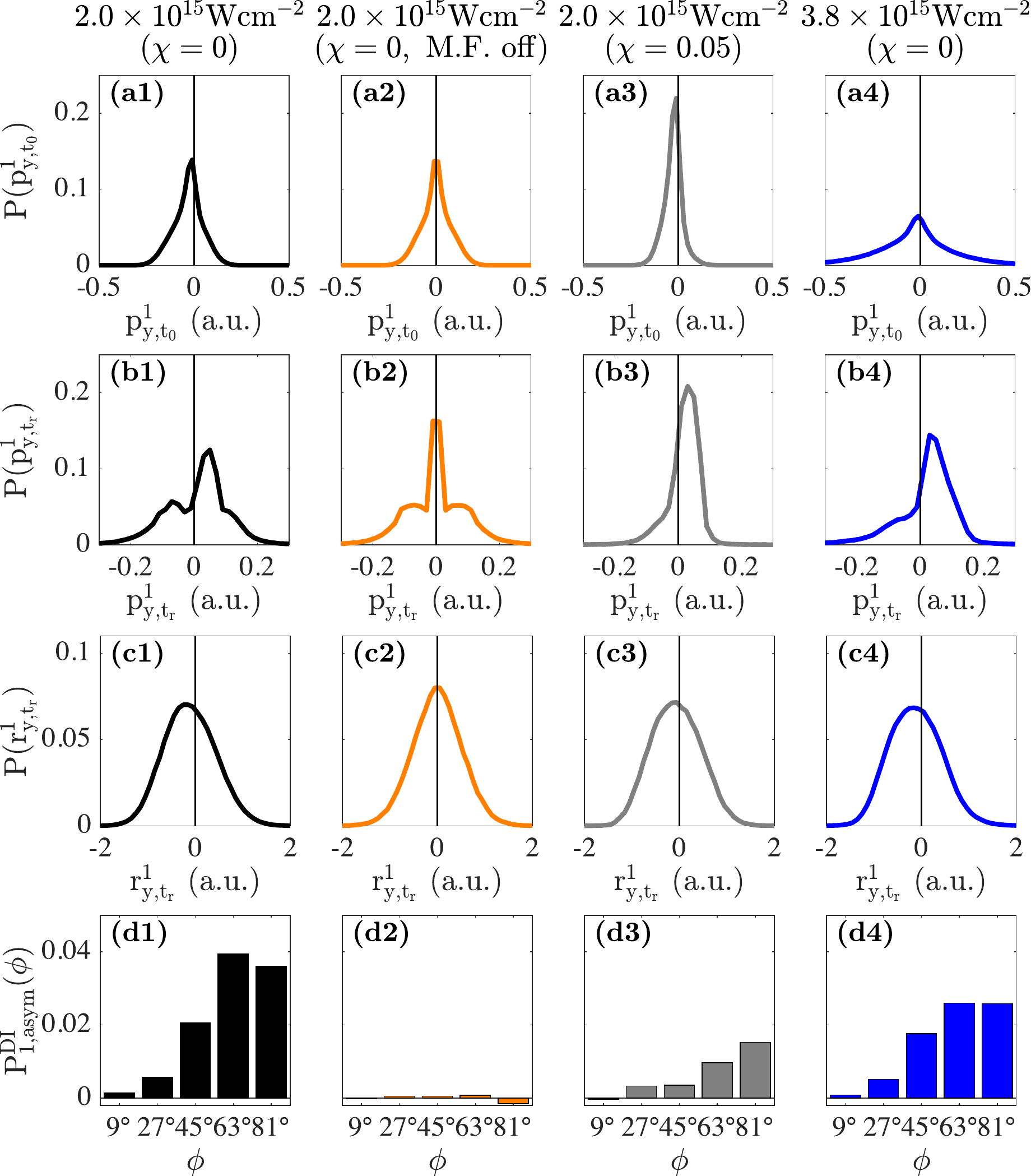}
\caption{In double ionization, the distributions of the y-component of the  electron 1 momentum at two different times are plotted; (a1)-(a4) at the time electron 1 tunnel-ionizes (momentum $\mathrm{p^1_{y,t_0}}$); (b1)-(b4) at the time  just before the time of recollision (momentum $\mathrm{p^1_{y,t_r}}$).The distributions of the y-component of the  electron 1 position  are plotted (c1)-(c4) at the time  just before the time of recollision (position $\mathrm{r^1_{y,t_r}}$).   $\mathrm{P^{DI}_{1,asym}(\phi)}$ is plotted in panels (d1)-(d4) for better comparison of its features with the features of $\mathrm{P(p^{1}_{y,t_r})}$. Panels (d1), (d3) and (d4) are the same as the plots in \fig{figure4}(b).  } 
\label{figure6}
\end{figure}

In the following sections we discuss the mechanism responsible for the features of P$\mathrm{_{1,asym}^{DI}(\phi)}$ and therefore for the features of the observable $\mathrm{P_{asym}^{DI}(\phi)}$.
We find that this mechanism is a  signature of recollision  exclusive  to non-dipole effects.  We adopt the term
 {\it non-dipole recollision-gated ionization} to describe it. 
We find that the magnetic field and the recollision  act together as a gate that selects only a subset of transverse initial momenta of the tunneling electron
that lead to double ionization. For linearly polarized light, this gating is illustrated in \fig{figure6}(a1) and (a2)  at an intensity of 2$\times$10$^{15}$ Wcm$^{-2}$ with the magnetic field switched-on and off, respectively.  We plot the probability distribution  $\mathrm{P(p^{1}_{y,t_0})}$ of electron 1  to tunnel-ionize  with a  y-component of the initial momentum  equal to $\mathrm{p_{y,t_0}^1}$.  We find that $\mathrm{P(p^{1}_{y,t_0})}$  is asymmetric when the magnetic field is switched-on. Specifically, it is  more likely for electron 1 to tunnel-ionize   with a negative rather than a positive  y-component of the initial momentum.  In addition, $\mathrm{P(p^{1}_{y,t_0})}$  peaks around small negative values of the momentum of electron 1. Instead,  $\mathrm{P(p^{1}_{y,t_0})}$ is symmetric when the magnetic field is switched-off.    
We also find that $\mathrm{P(p^{1}_{z,t_0})}$ is symmetric around zero (not shown) when the magnetic field is switched-on and off. This is expected since there is no force acting along the z-axis due to the laser field.  

We find that this asymmetry  in $\mathrm{P(p^{1}_{y,t_0})}$ persists at  a higher intensity  of 3.8$\times$10$^{15}$ Wcm$^{-2}$ for linearly polarized light, see \fig{figure6}(a4). We   find that an asymmetry in $\mathrm{P(p^{1}_{y,t_0})}$ is also present at  2$\times$10$^{15}$ Wcm$^{-2}$ for a laser pulse with a small  ellipticity of $\mathrm{\chi=0.05}$, see \fig{figure6}(a3).  In contrast, for all the above cases, we find that the initially bound electron 
has a symmetric distribution  $\mathrm{P(p^{2}_{y,t_0})}$ at the time electron 1 tunnel-ionizes and just before the time of recollision. Moreover, in single ionization we find that the escaping electron has a symmetric distribution  $\mathrm{P(p_{y,t_0})}$ at the time this electron tunnel-ionizes.

\subsection{Asymmetric transverse electron 1 momentum  shortly before recollision   }

In single ionization, to understand the features of $\mathrm{P_{asym}^{SI}(\phi)}$ one must obtain the distribution $\mathrm{P(p^{1}_{y,t_{0}})}$ at the time electron 1 tunnel-ionizes. Indeed, we compute  the y-component of the escaping electron's momentum    both with all Coulomb forces switched-off  and with all Coulomb forces accounted for.  In both cases we start the time propagation from the instant electron 1   tunnel-ionizes. We have shown in ref.\cite{Emmanouilidoumagnetic1} that the average  y-component of the electron 1 momentum  is roughly the same for both cases. At the time electron 1 tunnel-ionizes, we find that the y-component of the 
transverse momentum of electron 1 is roughly symmetric around zero. This initial momentum distribution combined with the   $\mathrm{{\bf F_{B}}}$ force give rise to the electron ionizing mostly with $\mathrm{\phi}$ slightly less than $\mathrm{90^{\circ}}$ or equivalently give rise to a sharply 
peaked distribution  P$\mathrm{^{SI}_{asym}(\phi)}$  (\fig{figure4}(f)).   

In double ionization, to understand the features of $\mathrm{P_{1,asym}^{DI}(\phi)}$, first, we must obtain the distribution $\mathrm{P(p^{1}_{y,t_{r}})}$ of the y-component of the electron 1 momentum  shortly before the time of recollision.   Then, we must find the effect of the recollision itself on  the distribution $\mathrm{P(p^{1}_{y,t_{r}})}$. The validity of these two steps  is supported by the following computations. We propagate the y-component of the momentum of electrons 1 and 2 from the time electron 1 tunnel-ionizes  up to the time of  recollision. We do so using the full 3D-SMND model  with all Coulomb forces accounted for.  Next, using as  initial conditions the momenta of electrons 1 and 2 
shortly after the time of recollision, we propagate from the time of recollision onwards with all Coulomb forces and the magnetic field switched-off. The final average y-component of the momentum of electron 1 is roughly equal in both cases  and the same holds for electron 2 \cite{Emmanouilidoumagnetic1}.
Therefore, the decisive time in double ionization is the time of recollision.  

Given the above, we first compute  $\mathrm{P(p^{1}_{y,t_{r}})}$ just before the time of recollision.  This is done by  extracting from the full 3D-SMND model  the distribution of the y-component of the electron 1 momentum  at a time just before the recollision, for instance at t$\mathrm{_{bef}=t_{r}-1/50T}$. T is the period of the laser field. We choose $\mathrm{t_{bef}}$  such as to avoid the  sharp change of the momenta which occurs at $\mathrm{t_{r}}$, see ref. \cite{Emmanouilidoumagnetic1}. 
At all intensities considered, we find that shortly before  the time of recollision a positive over a negative y-component of  electron 1 momentum  is favored. This is shown at intensities of 2$\times$10$^{15}$ Wcm$^{-2}$ and 3.8$\times$10$^{15}$ Wcm$^{-2}$ for linearly polarized light in \fig{figure6}(b1) and (b4), respectively, and at  2$\times$10$^{15}$ Wcm$^{-2}$ for elliptically polarized light with $\mathrm{\chi=0.05}$ in \fig{figure6}(b3). In contrast, when the magnetic field is switched-off the distribution  $\mathrm{P(p^{1}_{y,t_{r}})}$ is symmetric with respect to zero as illustrated for an intensity of 2$\times$10$^{15}$ Wcm$^{-2}$ in \fig{figure6}(b2). 

We have now established that the shift towards negative momenta of  $\mathrm{P(p^{1}_{y,t_{0}})}$ at the  time electron 1 tunnel-ionizes maps to a shift towards positive momenta of  $\mathrm{P(p^{1}_{y,t_{r}})}$ just before the time of recollision.
The width is another  interesting feature of the  distribution $\mathrm{P(p^{1}_{y,t_{0}})}$   and, by extension, of the distribution $\mathrm{P(p^{1}_{y,t_{r}})}$. 
From \fig{figure6}(b1)-(b4), we find that the width of $\mathrm{P(p^{1}_{y,t_{r}})}$ at 2$\times$10$^{15}$ Wcm$^{-2}$ with linearly  polarized light is the largest one, while the width of $\mathrm{P(p^{1}_{y,t_{r}})}$ at 2$\times$10$^{15}$ Wcm$^{-2}$ with $\mathrm{\chi=0.05}$ is the smallest one.
 Moreover, the width of $\mathrm{P(p^{1}_{y,t_{0}})}$ is comparable with the width of $\mathrm{P(p^{1}_{y,t_{r}})}$  for both the linear and the elliptical laser field at 2$\times$10$^{15}$ Wcm$^{-2}$. However,  the width of $\mathrm{P(p^{1}_{y,t_{0}})}$ is larger than the width of $\mathrm{P(p^{1}_{y,t_{r}})}$ at an intensity of 3.8$\times$10$^{15}$ Wcm$^{-2}$. 

We find that the  widths of $\mathrm{P(p^{1}_{y,t_{0}})}$ and $\mathrm{P(p^{1}_{y,t_{r}})}$ are consistent with Coulomb focusing which mainly refers to multiple returns of electron 1  to the core \cite{Coulomb1, Coulomb2}. In addition, for the larger  intensity of 3.8$\times$10$^{15}$ Wcm$^{-2}$, the widths are also  consistent with a larger effect of  the Coulomb potential of the ion on electron 1. The latter effect is not due to 
multiple returns of electron 1 to the core but rather due to electron 1 tunnel-ionizing closer to the nucleus at higher intensities. 

Indeed, we compute in Table I the number of times electron 1 returns to the core before it finally escapes. We find that electron 1 returns more times to the core at  2$\times$10$^{15}$ Wcm$^{-2}$. Namely, electron 1 escapes with only one   return to the core in 27\% of doubly-ionized events. Also, it returns roughly the same number of times at 2$\times$10$^{15}$ Wcm$^{-2}$ with $\mathrm{\chi=0.05}$ and at  3.8$\times$10$^{15}$ Wcm$^{-2}$. In these two latter cases, electron 1 escapes with only one return to the core in more than 50\% of doubly-ionized events.  We also find (not shown) that the width of the distributions $\mathrm{P(p^{1}_{y,t_{0}})}$ and $\mathrm{P(p^{1}_{y,t_{r}})}$ increases with  increasing number of returns to the core for all laser fields considered. The above features are consistent with Coulomb focusing. 
 Thus, Coulomb focusing explains why the width of  $\mathrm{P(p^{1}_{y,t_{0}})}$ at 2$\times$10$^{15}$ Wcm$^{-2}$ is larger than the width of   $\mathrm{P(p^{1}_{y,t_{0}})}$ at 2$\times$10$^{15}$ Wcm$^{-2}$ with $\mathrm{\chi=0.05}$.

\begin{table}[h]
\centering
\begin{ruledtabular}
\begin{tabular}{ccccc}
 & $\mathrm{2.0^\dagger}$ & $\mathrm{2.0^\dagger}$ & $\mathrm{2.0^\dagger}$ & $\mathrm{3.8^\dagger}$ \\
 & $\mathrm{(\chi=0)}$ & $\mathrm{(\chi=0, \ M.F. \ off)}$ & $\mathrm{(\chi=0.05)}$ & $\mathrm{(\chi=0)}$ \\
 \hline
$\hphantom{>}$1 Return$\hphantom{s}$ & 27\% & 24\% & 53\% & 57\% \\
$\hphantom{>}$2 Returns & 24\% & 24\% & 10\% & 16\% \\
$\hphantom{>}$3 Returns & 27\% & 27\% & 19\% & 14\% \\
$>$3 Returns & 21\% & 25\% & 18\% & 12\%
\end{tabular}
\end{ruledtabular}
$\mathrm{^\dagger}$ Intensities given in units of $10^{15}$ Wcm$^{-2}$
\caption{Number of returns to the core of electron 1 for doubly-ionized events.}
\end{table}
 Moreover, at the larger intensity of  3.8$\times$10$^{15}$ Wcm$^{-2}$ electron 1 exits the field-lowered Coulomb barrier closer to the nucleus. Indeed, we find that the average distance of electron 1 from the nucleus at the time electron 1 tunnel-ionizes is  2.4 a.u. at  3.8$\times
$10$^{15}$ Wcm$^{-2}$ compared to roughly 3.5 a.u.  at  2$\times$10$^{15}$ Wcm$^{-2}$ with $\mathrm{\chi=0}$ and $\mathrm{\chi=0.05}$, see Table II. This is consistent with the width of   $\mathrm{P(p^{1}_{y,t_{r}})}$ being significantly  smaller than the width of $\mathrm{P(p^{1}_{y,t_{0}})}$ at  3.8$\times$10$^{15}$ Wcm$^{-2}$. That is, the Coulomb potential of the ion has a large effect on electron 1 from the time electron 1 tunnel-ionizes onwards. This is not the case for the smaller intensities  of 2$\times$10$^{15}$ Wcm$^{-2}$ with $\mathrm{\chi=0}$ and $\mathrm{\chi=0.05}$ where the widths of the distributions  $\mathrm{P(p^{1}_{y,t_{0}})}$  and $\mathrm{P(p^{1}_{y,t_{r}})}$ are similar. 

 \begin{table}[h]
\centering
\begin{ruledtabular}
\begin{tabular}{ccccc}
 & $\mathrm{2.0^\dagger}$ & $\mathrm{2.0^\dagger}$ & $\mathrm{2.0^\dagger}$ & $\mathrm{3.8^\dagger}$ \\
 & $\mathrm{(\chi=0)}$ & $\mathrm{(\chi=0, \ M.F. \ off)}$ & $\mathrm{(\chi=0.05)}$ & $\mathrm{(\chi=0)}$ \\
 \hline
$\mathrm{\left<r^1_{t_0}\right>}$ (a.u.) & 3.5 & 3.5 & 3.7 & 2.4
\end{tabular}
\end{ruledtabular}
$\mathrm{^\dagger}$ Intensities given in units of $10^{15}$ Wcm$^{-2}$
\caption{Average distance from the nucleus of electron 1 at the time electron 1  tunnel-ionizes, $\mathrm{\left< r^1_{t_0}\right>}$.}
\end{table}

\subsection{Glancing angles  in recollisions }
We have shown in the previous section that effectively the only force that could result in a change of  the momenta of the two electrons between the final time and the time shortly after recollision is due to the electric field. Between the time shortly after recollision takes place and the asymptotic time, the electric field mainly affects the  x-component and not the y-component of the momentum of electron 1. Moreover, at the smaller intensities considered,  in the time interval following recollision, the electric field  does not affect significantly  the magnitude of the x-component of the electron 1 momentum; this component is mainly determined by the vector potential at the recollision time. 
Given the above, it is enough to find how the momentum of electron 1 (roughly equal to the x-component) and its y-component change from just before to just after the time of recollision due to the recollision itself. This will allow us to understand the features of the distribution  $\mathrm{P_{1,asym}^{DI}(\phi)}$ of the final angle $\mathrm{\phi}$. 

A measure of the strength of a recollision is   the angle $\mathrm{\theta}$, where $\mathrm{\cos{\theta}=p_{1,bef}\cdot p_{1,aft}/(|p_{1,bef}||p_{1,aft}|)}$. That is, $\mathrm{\theta}$, is the angle between the momentum of electron 1 just before and just after the time of recollision.  The momentum just before the  time of recollision  is roughly along the x-axis (polarization axis). Thus,  $\mathrm{\theta}$ is the angle of the momentum of electron 1  after the time of recollision with respect to  the x-axis. $\mathrm{\theta=180^{\circ}}$ corresponds to a ``head on" collision and complete 
 backscattering.  $\mathrm{\theta=0^{\circ}}$ corresponds to forward scattering and thus to almost no change due to the recollision.
 
\begin{figure} [ht]
\includegraphics[clip,height=0.4\textwidth]{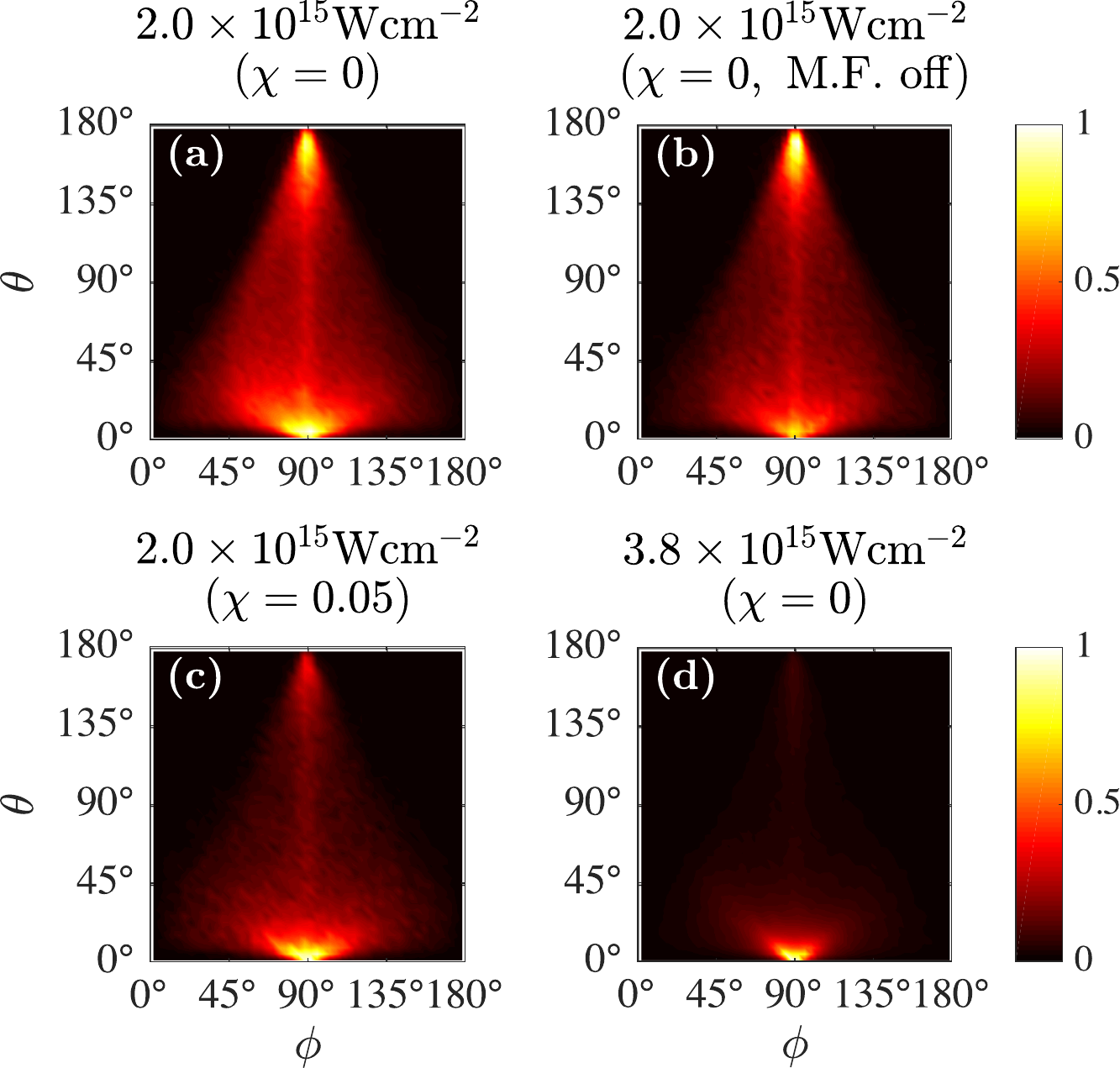}
\caption{Double differential probability of electron 1 to have a scattering angle $\mathrm{\theta}$ and a final angle $\mathrm{\phi}$ at  (a) 2$\times$10$^{15}$ Wcm$^{-2}$, (b)  2$\times$10$^{15}$ Wcm$^{-2}$ with the magnetic field switched-off, (c) 2$\times$10$^{15}$ Wcm$^{-2}$ with $\mathrm{\chi=0.05}$ and  (d) 3.8$\times$10$^{15}$ Wcm$^{-2}$.}
\label{figure5}
\end{figure}

  In \fig{figure5}, we show  for each laser field considered in this work, what is the probability for electron 1 to escape with a final angle $\mathrm{\phi}$  and a scattering angle  $\mathrm{\theta}$. It is shown that very strong recollisions, i.e. $\mathrm{\theta=180^{\circ}}$, take place only when electron 1 escapes with a momentum that has a very small y-component, i.e. $\mathrm{\phi}$ is around $\mathrm{90^{\circ}}$. However, even when electron 1 escapes  with $\mathrm{\phi}$ around $\mathrm{90^{\circ}}$,  it is more likely that a weak recollision takes place, i.e.  $\mathrm{\theta=0^{\circ}}$, rather than a strong one with $\mathrm{\theta=180^{\circ}}$.   Moreover, when electron 1 ionizes with momenta that have larger y-components with $\mathrm{\phi\in [45^{\circ},90^{\circ}]}$ and $\mathrm{\phi\in [90^{\circ},135^{\circ}]}$ the scattering angles $\mathrm{\theta}$ are on average smaller than $\mathrm{90^{\circ}}$. That is, in most cases, electron 1 ionizes at  glancing angles $\mathrm{\theta}$ following recollision. Moreover, a comparison of the values of $\mathrm{\theta}$ at 2$\times$10$^{15}$ Wcm$^{-2}$ and at 3.8$\times$10$^{15}$ Wcm$^{-2}$, see \fig{figure5}(a) and (d), clearly shows that overall the recollision is stronger at the smaller intensity.  In addition, a comparison  of the values of $\mathrm{\theta}$ at 2$\times$10$^{15}$ Wcm$^{-2}$ and at 2$\times$10$^{15}$ Wcm$^{-2}$ with $\mathrm{\chi=0.05}$, see \fig{figure5}(a) and (c), clearly shows that overall the recollision is stronger when $\mathrm{\chi=0}$.  Thus,  electron 1 escapes at glancing angles following recollision. 
 
\subsection{Asymmetric transverse electron 1 position shortly before recollision} 
We first explain how the y-component of the momentum of electron 1  changes from just before to just after the time of recollision, due to the recollision itself, when the magnetic field is switched-off at 2$\times$10$^{15}$ Wcm$^{-2}$.  We find that  doubly-ionized events are equally likely to have positive or negative y-component of the momentum and of the position of electron 1 just before the time of recollision, see \fig{figure6}(b2) an (c2). Moreover, we find that the y-component of the final momentum of electron 1  is positive (negative) depending on whether the y-component of the position of electron 1 is negative (positive) just before the time of recollision. The reason is that electron 1 ionizes at glancing angles.  The direction of the Coulomb attraction of electron 1 from the nucleus just before the time of recollision   determines whether  just after the time of recollision, as well as at asymptotically large times,  the y-component of the momentum of electron 1 is positive or negative. Indeed, we find that doubly-ionized events with  $\mathrm{r^{1}
_{y,t_r}>0}$ give rise to negative values of $\mathrm{P^{DI}_{1,asym}}$, while doubly-ionized events with  $\mathrm{r^{1}
_{y,t_r}<0}$ give rise to positive values of   $\mathrm{P^{DI}_{1,asym}}$ and cancel each other out.

However, when the magnetic field is switched-on at 2$\times$10$^{15}$ Wcm$^{-2}$,  most (59\%) doubly-ionized events have $\mathrm{r^{1}_{y,t_r}<0}$ just before the time of recollision. This is due to  {\it non-dipole recollision gated ionization}, since out of these latter events 61\% have   both $\mathrm{p^{1}_{y,t_r}>0}$ just before the time of recollision and $\mathrm{p^{1}_{y,t_0}<0}$ at the initial time that electron 1 tunnel-ionizes. The shift towards negative values  in  the transverse y-component of the position  of electron 1 just before the time of recollision is shown  in \fig{figure6}(c1), (c3) at 2$\times$10$^{15}$ Wcm$^{-2}$ 
with $\mathrm{\chi=0}$ and $\mathrm{\chi=0.05}$, respectively,  and in (c4) at 3.8$\times$10$^{15}$ Wcm$^{-2}$.  As for the case when the magnetic field is switched-off,  when the magnetic field is switched-on  electron 1 ionizes at glancing angles. Therefore,   we find that doubly-ionized events with  $\mathrm{r^{1}
_{y,t_r}>0}$ give rise to negative values of $\mathrm{P^{DI}_{1,asym}}$ and doubly-ionized events with  $\mathrm{r^{1}
_{y,t_r}<0}$ give rise to positive values of   $\mathrm{P^{DI}_{1,asym}}$. However, the latter events are 59\% of all doubly-ionized events and thus overall $\mathrm{P^{DI}_{1,asym}}$ has  positive values.

\begin{figure} [ht]
\includegraphics[clip,width=0.45\textwidth]{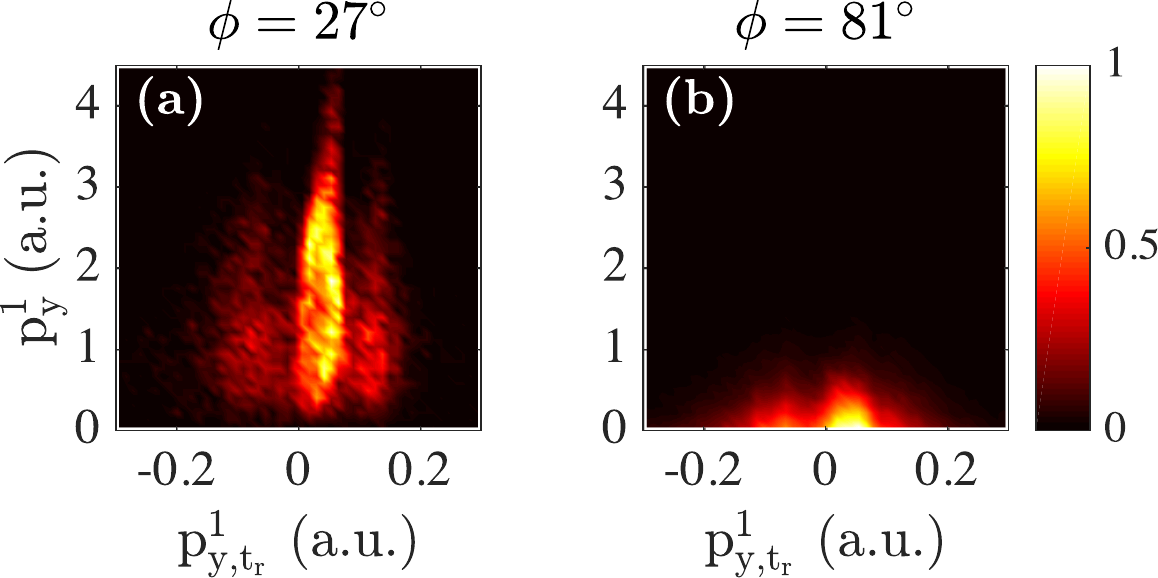}
\caption{Double differential probability of the tunneling electron to have a y-component of the  momentum equal to $\mathrm{p^{1}_{y,t_r}}$  at the time of recollision and equal to $\mathrm{p^{1}_{y}}$ at the asymptotic time for  two different $\mathrm{\phi}$s  at $\mathrm{2.0\times10^{15}W/cm^2}$. $\mathrm{\phi}$ is binned in intervals of 18$^{\circ}$.} 
\label{figure7}
\end{figure}

From the above it also follows  that a small width of $\mathrm{P(p^{1}_{y,t_{r}})}$ affects only doubly-ionized events with small    y-components   of the electron 1  final momenta. On the other hand, a large  width of $\mathrm{P(p^{1}_{y,t_{r}})}$  just before the time of recollision affects doubly-ionized events with     y-components   of the electron 1  final momenta ranging from small to large. However,   doubly-ionized events with large y-components of the electron 1 momenta correspond to smaller $\mathrm{\phi}$ in  $\mathrm{P^{DI}_{1,asym}}$.
  Indeed,  this is clear  in \fig{figure7} when comparing the distribution of the y-component of the final momentum of  electron 1 in the smaller $\mathrm{\phi}$ of $\mathrm{27^{\circ}}$ with the one in the larger  $\mathrm{\phi}$ of $\mathrm{81^{\circ}}$.   The large y-components  of the momenta of electron 1  are a result of the recollision. This is the case since   for each intensity,  $\mathrm{P(p^{1}_{y,t_{r}})}$ are similar for all  $\mathrm{\phi}$ bins, see \fig{figure7}.  $\mathrm{P(p^{1}_{y,t_{r}})}$  is depicted in \fig{figure6}(b1), (b3) and (b4) for three different laser fields. 
  Most doubly-ionized events just before the time of recollision have positive y-component of the momentum of electron 1. Thus, the width of 
 $\mathrm{P(p^{1}_{y,t_{r}})}$ is also a measure of the shift towards positive y-components of the final momenta of electron 1.  
 
 Finally, we compare the width of the distribution of the x-component of the momentum of electron 1  at intensities where strong recollisions prevail  with  the width at intensities where soft ones do.  We find that this width is larger just before and just after the time of recollision as well as at asymptotically large times at intensities where soft recollisions prevail. Indeed, for strong recollisions, the x-component of the momentum of electron 1 is determined mainly by the vector potential at the time of recollision. However, for soft recollisons this is not quite the case since electron 1 only transfers a small amount of its energy to electron 2. In addition, for soft recollisions, the  time of recollison has a much broader range of values, see ref.\cite{Emmanouilidoumagnetic1}. 

We now combine the width of the asymmetry of $\mathrm{P(p^{1}_{y})}$ with the width  of the distribution of the x-component of the momentum of electron 1.  It then follows that  
  $\mathrm{P_{1,asym}^{DI}(\phi)}$ should have  higher values over a larger range of  $\mathrm{\phi}$ at 2$\times$10$^{15}$ Wcm$^{-2}$ (stronger recollision) and  smaller values at 2$\times$10$^{15}$ Wcm$^{-2}$ with $\mathrm{\chi=0.05}$ (softer recollision). Indeed, this is the case as shown in \fig{figure6}(d1)-(d4).

 \section{Average sum electron momenta in double ionization}

\begin{figure} [h]
\includegraphics[clip,height=0.22\textwidth]{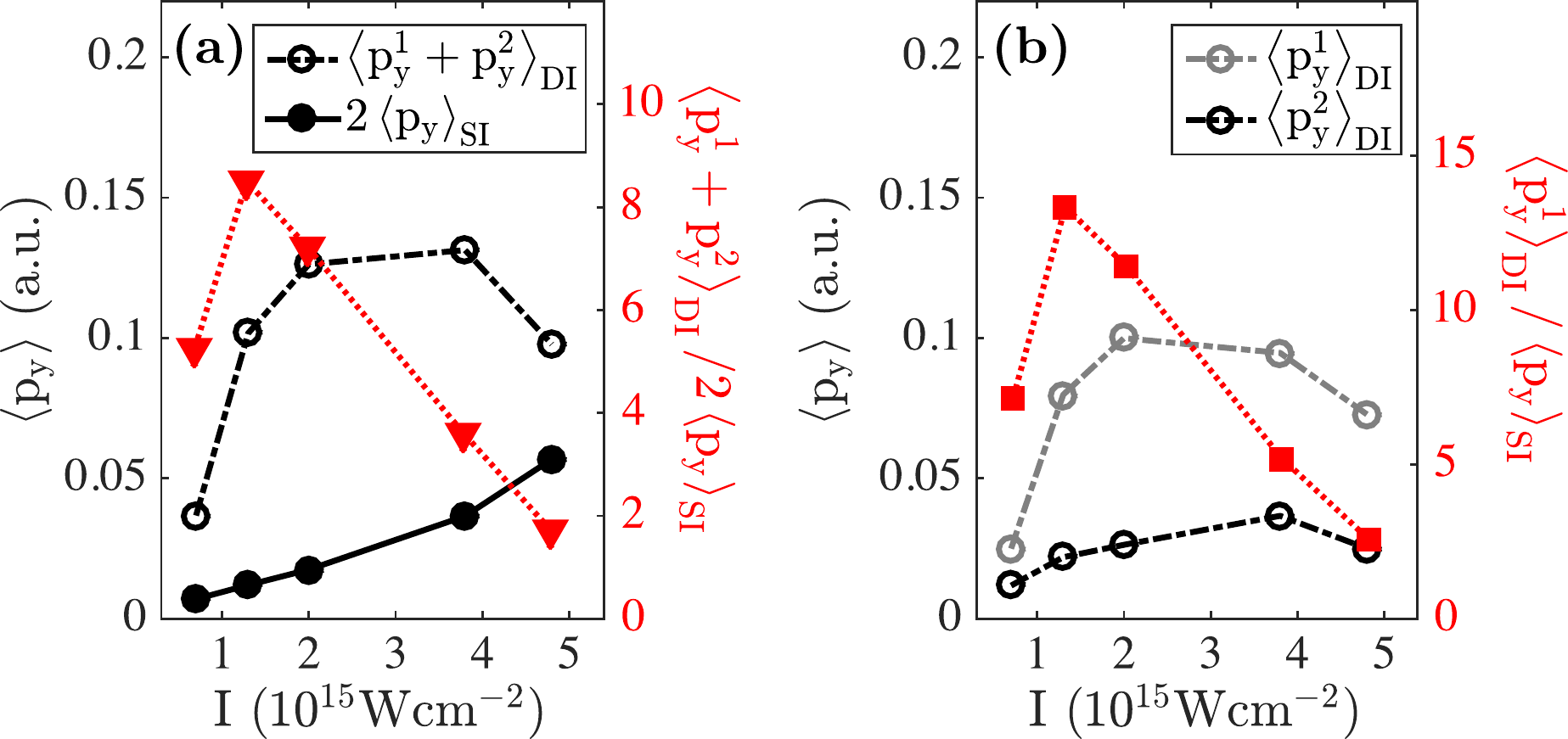}
\caption{(a) The average sum of the two electron momenta $\mathrm{\left <p_{y}^1+p_{y}^{2} \right >_{DI}}$ in double ionization (black dot-dashed line with open circles), twice the average electron momentum $\mathrm{\left <p_{y}\right>_{SI}}$ in single ionization (black solid line with circles) and the ratio $\mathrm{\left <p_{y}^1+p_{y}^{2} \right >_{DI}/\left <p_{y}\right>_{SI}}$ (red dotted line with triangles) as a function of the intensity of the laser field. (b) The average momentum of the tunneling electron $\mathrm{\left <p^{1}_{y}\right>_{DI}}$ (grey dot-dashed line with open circles)  and the bound electron $\mathrm{\left <p^{2}_{y}\right>_{DI}}$ (black dot-dashed line with circles) in double ionization and the ratio $\mathrm{\left <p_{y}^1 \right >_{DI}/\left <p_{y}\right>_{SI}}$ (red dotted line with squares) as a function of the intensity of the laser field. } 
\label{figure1}
\end{figure}

 In ref.\cite{Emmanouilidoumagnetic1}, we have shown that in double ionization the ratio $\mathrm{\left <p_{y}^1+p_{y}^{2} \right >_{DI}/2\left <p_{y}\right>_{SI}}$ is maximum and roughly equal to eight at intensities 1.3$\times$10$^{15}$ Wcm$^{-2}$ and 2$\times$10$^{15}$ Wcm$^{-2}$,  see \fig{figure1}. $\mathrm{\left <p_{y}^1+p_{y}^{2} \right >}$ is the 
 average sum of the two electron momenta along the propagation direction of the laser field, while $\mathrm{\left <p_{y}\right>_{SI}}$
is the corresponding average  electron momentum in single ionization. In \fig{figure1}(a),  $\mathrm{\left <p_{y}^1+p_{y}^{2} \right >_{DI}/2\left <p_{y}\right>_{SI}}$ is shown to decrease  with increasing intensity, for the intensities considered.  Moreover, in \fig{figure1}(b), it is  shown that it is $\mathrm{\left <p_{y}^1\right>_{DI}}$ of the tunneling electron that contributes the most to  $\mathrm{\left <p_{y}^1+p_{y}^{2} \right >_{DI}}$ for the intensities considered. 
The ratio   $\mathrm{\left <p_{y}^1 \right >_{DI}/\left <p_{y}\right>_{SI}}$ has surprisingly    large values  at intensities smaller than the intensities satisfying  the criterion for the onset of magnetic-field effects $\mathrm{\beta_{0}\approx}$1 a.u.  \cite{Reiss1, Reiss2}.
    In contrast, we find that $\mathrm{\left <p_{y}\right>_{SI}}$ increases from 0.0035 a.u.  at 0.7$\times$10$^{15}$ Wcm$^{-2}$ to 0.028 a.u. at 4.8$\times$10$^{15}$ Wcm$^{-2}$, see \fig{figure1}(a) and  ref.\cite{Emmanouilidoumagnetic1}. The small values of the average  electron momentum in single ionization and the increase of this average with intensity are in accord with the effect of the  $\mathrm{\bf F_{B}}$ force.  The latter increases with intensity,  since the magnetic field increases.   The increase of  $\mathrm{\left <p_{y}\right>_{SI}}$ with intensity has been addressed in  several experimental and theoretical studies \cite{Corkum1, Corkum2,Corkum3, Drake, IvanovA}.

\begin{figure} [ht]
\includegraphics[clip,height=0.15\textwidth]{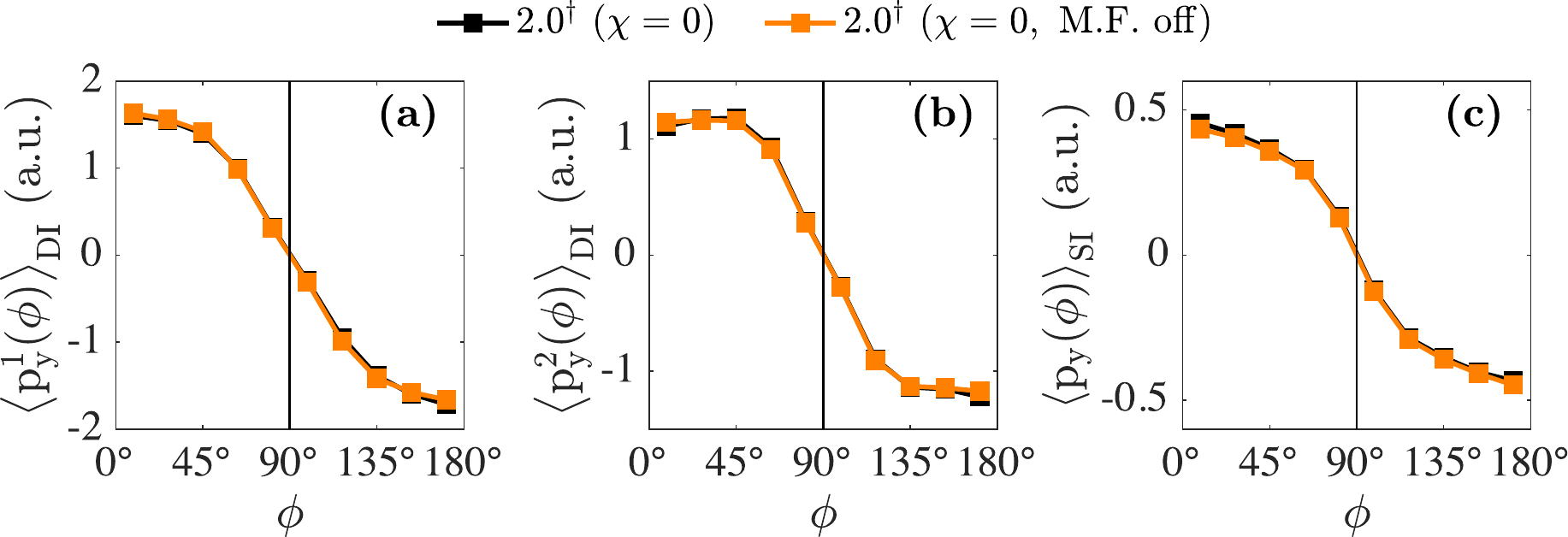}
\caption{(a) $\mathrm{\left <p_{y}^{1}\right>_{DI}}$  of  the tunneling electron,  (b) $\mathrm{\left <p_{y}^{2}\right>_{DI}}$  of the initially bound electron and (c) $\mathrm{\left <p_{y}\right>_{SI}}$ in single ionization are plotted as a function of $\mathrm{\phi}$ at  2$\times$10$^{15}$ Wcm$^{-2}$ with the magnetic field  switched-on and off.  $\mathrm{\phi}$ is binned in intervals of 18$^{\circ}$. $\dagger$ denotes a multiplication factor of $\mathrm{10^{15}Wcm^{-2}}$. }
\label{figure2}
\end{figure}

In what follows, we show that {\it non dipole recollision gated ionization } accounts for the large values of $\mathrm{\left <p_{y}^1 \right >_{DI}/\left <p_{y}\right>_{SI}}$ and thus for the large average sum of the two electron momenta along the propagation direction of the laser field at smaller intensities of 1.3-2$\times$10$^{15}$ Wcm$^{-2}$. To do so, we express $\mathrm{\left <p_{y}^i\right>_{DI}}$  as 
  \begin{equation}  
  \mathrm{\left <p_{y}^i\right>_{DI}}= \mathrm{\int_{0^{\circ}}^{180^{\circ}} \left <p_{y}^i(\phi)\right >_{DI}P_{i}^{DI}(\phi)}d\phi,
\label{eq1}
\end{equation}  
with $\mathrm{i=1,2}$ for electrons 1 and 2 in double ionization. A similar expression holds in single ionization.   We have already shown  that {\it non-dipole recollision gated ionization}  accounts for the asymmetry  in $\mathrm{P_{1}^{DI}(\phi)}$. 
      Next, we investigate the influence of the magnetic field on $\mathrm{\left <p^{i}_{y}(\phi)\right>_{DI}}$. 
   To do so, $\mathrm{\left <p_{y}^1(\phi)\right>_{DI}}$  of the tunnel electron, $\mathrm{\left <p^{2}_{y}(\phi)\right>_{DI}}$ of the bound electron  and $\mathrm{\left <p_{y}(\phi)\right>_{SI}}$  are plotted in \fig{figure2}(a), (b) and (c), respectively, at  2$\times$10$^{15}$ Wcm$^{-2}$  with linearly polarized light and with the magnetic field switched-on and off. It is shown that the magnitude of the average electron momentum increases as a function of $\mathrm{\phi}$, both in double and in single ionization.
  This is evident mostly for $\mathrm{\phi}$ around 0$^{\circ}$ and 180$^{\circ}$. Moreover, it is clearly seen that the magnetic field has no influence on any of the average electron momenta considered and  that $\mathrm{\left <p_{y}^{i}(\phi)\right>_{DI}=\left<p_{y}^{i}(180^{\circ}-\phi)\right>_{DI}}$ and $\mathrm{\left <p_{y}(\phi)\right>_{SI}=\left<p_{y}(180^{\circ}-\phi)\right>_{SI}}$. This is expected when the magnetic field is switched-off, since there is no preferred direction of electron escape on the plane that is perpendicular to the polarization direction (x-axis) of the laser field. We find that similar results hold at 1.3$\times$10$^{15}$ Wcm$^{-2}$ and at 3.8$\times$10$^{15}$ Wcm$^{-2}$.

 We also find (not shown) that the effect of the magnetic field on $\mathrm{\left <p_{y}^1(\phi)\right >_{DI}}$ is very small  even at a more differential level. Specifically, the dependence of $\mathrm{\left <p_{y}^1(\phi,t)\right >_{DI}}$ on time is very similar both with the magnetic field switched-on and off.  The only difference is a small oscillation due to the magnetic field. It is noted that $\mathrm{\left <p_{y}^1(\phi,t\rightarrow \infty )\right >_{DI}=\left <p^{1}_{y}(\phi)\right>_{DI}}$. We find that similar results (not shown) hold for the bound electron.

 Using \eq{eq1}, we can now explain why $\mathrm{\left <p_{y}^{1}\right>_{DI}}$ is much larger than $\mathrm{\left <p_{y}\right>_{SI}}$ at intensities around 2$\times$10$^{15}$ Wcm$^{-2}$. When the magnetic field is switched-on, $\mathrm{\left <p_{y}^1(\phi)\right >_{DI}}$ does not change but
  P$\mathrm{^{DI}_{1,asym}(\phi)}$ does.
  $\mathrm{P_{1,asym}^{DI}(\phi)}$  is much wider than $\mathrm{P_{asym}^{SI}(\phi)}$ and than $\mathrm{P_{2,asym}^{DI}(\phi)}$ and has higher values at 2$\times$10$^{15}$ Wcm$^{-2}$ rather than at 3.8$\times$10$^{15}$ Wcm$^{-2}$. The  higher values of   $\mathrm{P_{1,asym}^{DI}(\phi)}$ over a wider range of $\mathrm{\phi}$ compared to P$\mathrm{^{SI}_{asym}(\phi)}$ and  $\mathrm{P_{2,asym}^{DI}(\phi)}$  result in smaller $\mathrm{\phi}$ and thus  larger  $\mathrm{\left <p_{y}^{1}(\phi)\right>_{DI}}$ (\fig{figure2}) having a significant weight  in   \eq{eq1}.

\section{Conclusions}
We account for  non-dipole effects in double ionization. We show that the recollision and the magnetic field act together as a gate. This gate gives rise to a distribution of the  y-component   of the tunneling electron initial momentum, $\mathrm{P(p^{1}_{y,t_{0}})}$, which is shifted towards negative values; negative values in the y-axis are opposite to the propagation direction of the laser field.  The term {\it non-dipole recollision-gated ionization} was adopted to describe this effect. We show that this asymmetry in $\mathrm{P(p^{1}_{y,t_{0}})}$ maps in time to an asymmetry of the transverse electron 1 momentum  just before the   time of recollision, i.e. to an asymmetry in $\mathrm{P(p^{1}_{y,t_{r}})}$. Namely, the y-component of the momentum of the tunneling electron is shifted towards positive values just before the time of recollision. 
Moreover, the asymmetry in $\mathrm{P(p^{1}_{y,t_{0}})}$ maps in time to an asymmetry of the transverse electron 1 position just before the time of recollision. Namely,   the y-component of the position  of the tunneling electron is shifted towards negative values just before the time of recollision. 

The above asymmetries combined with  the tunneling electron escaping at glancing angles following a recollision 
give rise  to an asymmetry in $\mathrm{P_{1}^{DI}(\phi)}$  with respect to $\mathrm{\phi=90^{\circ}}$. The latter is the probability distribution   of electron 1 to escape with an angle  $\mathrm{\phi}$. 
We find the asymmetry in $\mathrm{P_{1}^{DI}(\phi)}$ to be more significant, i.e. higher values of   $\mathrm{P_{1, asym}^{DI}(\phi)}$ over a  wider range of $\mathrm{\phi}$, in double ionization  compared to single ionization. Moreover, we find that higher values of   $\mathrm{P_{1, asym}^{DI}(\phi)}$ over a  wider range of $\mathrm{\phi}$ result 
from larger widths of $\mathrm{P(p^{1}_{y,t_{r}})}$ just before the time of recollision. The latter width depends on the intensity and the ellipticity of the laser pulse and we show that it is related to Coulomb focusing.   
We also show that it is the asymmetry in $\mathrm{P_{1}^{DI}(\phi)}$ over a wide range of $\mathrm{\phi}$ that accounts for the large values of the average transverse electron 1 momentum  and thus of the large average sum of the two electron  momenta at smaller  intensities.  Even though not as pronounced, we find these features of the probability distribution $\mathrm{P_{1}^{DI}(\phi)}$ of the tunneling electron  to also be present 
in an experimentally accessible observable. Namely, the probability  distribution for electron 1 or 2 to escape with an angle $\mathrm{\phi}$. This observable  effect of the {\it non-dipole recollision-gated ionization} can be measured by future experiments.

Finally, in the current work we show that  magnetic field effects  cause an offset of the  transverse momentum and position of the recolliding electron just before recollision takes place. We show how these asymmetries  lead to asymmetries in observables experimentally accessible. We conjecture that this is a general phenomenon not restricted to magnetic field effects. Namely, these observable asymmetries will be present  in any process that has two delayed steps and allows an electron to gain an offset before recollision takes place.

\section{Acknowledgments}A.E. acknowledges the EPSRC grant no. J0171831 and the use of the computational resources of Legion at UCL.

\end{document}